\documentclass[12pt]{article}
\begin{document}
\title{Mixture of Quantum States :\\ Thermal and Interaction Inducing Decoherence}
\author{{Kentaro Urasaki}%
\footnote{email address: urasaki@f6.dion.ne.jp}}
\date{}%

\maketitle

\abstract{
In this study, we show that the interaction energy plays an important role on the quantum decoherence:
If we pay attention to the oscillation phase factor, $e^{-iE_{int}t/\hbar},$ 
we see that the time average of the macro-system's density matrix becomes nearly diagonal, where the states giving extrema of interaction energy are privileged to describe the quantum decoherence. 
This approach is compatible with the von Neumann's old work, which has been recently studied with renewed interest: The thermal mixture of states can be reached by the time average of a density of matrix due to the oscillation phase factor, $e^{-i(E_i-E_j)t/\hbar}.$
One of the direct results is the localization of macroscopic objects.}

\vspace{2cm}

\section{Introduction}
The contorol of the quantum decoherence is recognized to be important when we make use of quantum technology. 
In such the situation, it is natural that now we are forced to reconsider the origin of our classical world itself. 
In fact, many researches in this area have been opening up the possibility to explain the origin of our classical world with the use of quantum mechanics in more integrated manner. 
In this study, we attempt to clarify the precise mechanism for the emergence of mixture of states 
resulting from the quantum decoherence. 

\bigskip

\subsection{Quantum thermal mixture}
In his 1929 paper\cite{Neumann1929}, von Neumann studied the scenario that the approximate mixture of states can be driven directly from quantum mechanics, where the time average of density matrices is crucially important. 
This mixture of states originates from the difference of the energy eigen-values and almost reproduce the micro-canonical ensemble. 
There have been a renewed interest in this approach.\cite{Goldstein2010}

\bigskip
The system described by the thermodynamics or the statistical mechanics has the finite entropy. 
This means that the system can be considered to be in the mixture of states despite the fact that a pure quantum state always has a zero entropy in its strict sense.
The simple solution to this problem is that the density matrix becomes almost diagonal when it is represented by energy eigenstates. 
Off-diagonal terms having the oscillating factor, $e^{i(E_i-E_j)t/\hbar}$, vanishes by a long time average (assuming $E_i\ne E_j$). 
Namely, the approximate mixture of states emerges, where the energy eigenstate is privileged. 
This is the simplest case that mixture of states coexists with the unitary time evolution of the quantum mechanics.

\subsection{Another mixture of states induced by the interaction}
On the other hand, there is another problem relevant to the mixture of quantum states, which is known as the quantum decoherence. 
The quantum decoherence can potentially explain the classical behavior of macroscopic quantum systems using its important concepts such as the existence of the pointer states, the localization of the object and so on. 

\bigskip

Also to this subject, von Neumann presented the starting point through the measurement problem and analyzed its process using a fully quantum mechanical model\cite{Neumann1932}:
He suggested that if we use the basis diagonalizing the interaction Hamiltonian in order to expand 
the state of a quantum system, each term resultingly corresponds to the state measured by us.
Let $|n_c\rangle$ denote the state of the system and $|\Phi_0\rangle$ denote its environment.
(In the context of our study, the both systems are assumed to be macroscopic. )
For example, in the case that the interaction Hamiltonian is dominant and is represented by the appropriate basis as, 
\begin{eqnarray}
\hat{h}_{int}=\sum_{n_c}|n_c\rangle\langle n_c|\otimes \hat{A}_{n_c},
\end{eqnarray}
then the state vector evolves as 
\begin{eqnarray}
|\Phi(0)\rangle=\sum_{n_c} |n_c\rangle|\Phi_0\rangle \to |\Phi(t)\rangle&&=\sum_{n_c} e^{-i\hat{h}_{int}t/\hbar}|n_c\rangle|\Phi_0\rangle\\
&&=\sum_{n_c} e^{-i\hat{A}_{n_c}t/\hbar}|n_c\rangle|\Phi_0\rangle,
\end{eqnarray}
(we use the Schr\"odinger representation).
In this case, it seems reasonable to consider that each state, $e^{-i\hat{A}_{n_c}t/\hbar}|n_c\rangle|\Phi_0\rangle$, corresponds to the result of the measurement.
However, the right hand side of this equation is still in the superposition of different $n$. 
Therefore, at least, it is necessary to derive the loss of the coherence between different $n$ and the resulting emergence of mixture of states. 
The decoherence scenario is an attempt to answer this problem.
\footnote{It is beyond the author's ability to introduce all the important researches on this area. You can extract more information from the citations of \cite{Joos1985, Schlosshauer2005, Zurek1981}.}

\bigskip

We here propose a new approach to this subject:
The analogue of the previous thermal mixture case is that taking into account both the time  average and the interaction energy naturally leads the emergence of the quantum decoherence.
Of course, there is, however, a problem with this simple but naive intuition:
Why is the interaction energy privileged in the phase factor, instead of the exact energy of a system?
The present study can answer this:
the states which diagonalize the interaction Hamiltonian cause the entanglement between the system and the environment, where we find the necessity for the dense spectrum.

\subsection{Coexistence of unitary evolution and decoherence}
In the case that the interaction Hamiltonian is non-commutative with the system's Hamiltonian, how do these two types of time evolution balance each other out?
Such the process is not a priori from the unitary evolution of quantum mechanics and the transitional behavior will be essential.
We will see below that it can potentially reproduce the dynamics often observed in our daily experience: one example is the localization of macroscopic objects.
(On the other hand, for the two-state system, we will also see that the conditions for the quantum decoherence cannot be satisfied when the interaction Hamiltonian is non-commutative with the system's Hamiltonian.)

\vspace{1cm}

\section{Interaction energy inducing decoherence}
Ordinary macroscopic systems weakly interact with its environment. 
As the result, the form or the position of these systems can be the origin of the difference in the interaction energy. 
Namely if we adopt the perturbation approach to the interaction, the branches will emerge in the energy level of the system.
For simplicity, we assume that initially the system has discrete energy levels and the interaction Hamiltonian induces a continuous spectrum disturbance, below.

\subsection{Emergence of mixture of states}
We use the time-dependent state vectors and start from the initial state in the simple product  form of normalized vectors, $|\Phi(t_0)\rangle=|\phi(t_0)\rangle|\varepsilon(t_0)\rangle$, where $|\phi(t_0)\rangle$ denotes the state of a macroscopic system and $|\varepsilon(t_0)\rangle$ denotes its environment. 
These non-perturbative states are defined to obey Schr\"odinger equations: 
\begin{eqnarray}
&&[i\hbar\partial_t-\hat{h}_\phi]|\phi(t)\rangle=0, \label{eq:subsystem1}\\
&&[i\hbar\partial_t-\hat{h}_\varepsilon]|\varepsilon(t)\rangle=0,\label{eq:subsystem2}
\end{eqnarray}
where $\hat{h}_\phi$ ($\hat{h}_\varepsilon$) is the Hamiltonian of the subsystem 
to act only on $|\phi(t)\rangle$ ($|\varepsilon(t)\rangle$).
Then, the non-perturbative state of the total system $|\Phi_0(t)\rangle=|\phi(t)\rangle|\varepsilon(0)\rangle$ also satisfies, 
\begin{eqnarray}\label{eq:whole0}
[i\hbar\partial_t-\hat{h}_\phi-\hat{h}_\varepsilon]|\Phi_0(t)\rangle=0. 
\end{eqnarray}

Let us consider the effect of the interaction between these systems, $\hat{h}_{int}$: 
\begin{eqnarray}\label{eq:whole}
[i\hbar\partial_t-\hat{h}_\phi-\hat{h}_\varepsilon-\hat{h}_{int}]|\Phi(t)\rangle=0. 
\end{eqnarray}
First, we do the calculation simply assuming that we can use the basis of the non-perturbative  system, $\{|\Phi_{0n}(t)\rangle\}$, which has completeness at each time $t$, to expand a state of the total system even when the interaction, $\hat{h}_{int}$, exists.
We can expand a perturbed state in the same way as the standard perturbation theory does, 
\begin{eqnarray}\label{eq:expansion}
|\Phi(t)\rangle=\sum_nC_n(t)|\Phi_{0n}(t)\rangle=\sum_nC_n(t)e^{-i(\hat{h}_\phi+\hat{h}_\varepsilon)(t-t_0)/\hbar}|\Phi_{0n}(t_0)\rangle.
\end{eqnarray}
The time dependence of the coefficient $C_n(t)$ originates from the interaction effect.
Substituting this into Eq. (\ref{eq:whole}) and acting $\langle\Phi_{0n}(t)|$, we obtain 
\begin{eqnarray}
i\hbar\partial_tC_n(t)=\sum_{n^\prime}C_{n^\prime}(t)\langle\Phi_{0n}(t)|\hat{h}_{int}|\Phi_{0n^\prime}(t)\rangle.
\end{eqnarray}
For weak interaction, the standard perturbation theory usually teaches us that $C_{n^\prime}(t)$'s in the right hand side can be replaced by $C_{n^\prime}(t_0)$'s under the assumption of the weak time dependence. 
The factor $\Lambda_n/\hbar$, however, becomes large for the contact of the macroscopic systems in the present case that we cannot neglect the time dependence of $C_{n^\prime}(t)$'s.

If there are sufficient number of states (see \S 4) or the states are well separated (see \S 3.1), we can choose the appropriate basis, $\{|\Phi_{0n}(t)\rangle\}$, in order to diagonalize the interaction Hamiltonian, $\hat{h}_{int}$, where $\langle\Phi_{0n}(t)|\hat{h}_{int}|\Phi_{0n^\prime}(t)\rangle\simeq\delta_{n,n^\prime}\langle\Phi_{0n}(t)|\hat{h}_{int}|\Phi_{0n}(t)\rangle.$
Therefore we obtain,
\begin{eqnarray}
i\hbar\partial_tC_n(t)\simeq C_n(t)\langle\Phi_{0n}(t)|\hat{h}_{int}|\Phi_{0n}(t)\rangle\label{eq:diagonal}.
\end{eqnarray}
We can easily integrate it as $C_n(t)=C_n(t_0)e^{-i\Lambda_n(t)/\hbar}$, where 
\begin{eqnarray}
\Lambda_n(t)=\int^t_{t_0}\langle\Phi_{0n}(t)|\hat{h}_{int}|\Phi_{0n}(t)\rangle dt
\end{eqnarray} 
and, 
\begin{eqnarray}\label{eq:expansion2}
|\Phi(t)\rangle\simeq\sum_nC_n(t_0)|\Phi_{0n}(t)\rangle e^{-i\Lambda_n(t)/\hbar}.
\end{eqnarray}
Therefore if the initial state is in the state, $|\Phi(t_0)\rangle=|\Phi_{0n}(t_0)\rangle,$ this state evolves into $|\Phi_n(t)\rangle=|\Phi_{0n}(t)\rangle e^{-i\Lambda_n(t)/\hbar}$, 
where only the oscillational phase factor due to the interaction energy is taken into account.  

Since the interaction energy is sufficiently large in the present case, the off-diagonal elements in the density matrix represented by the present basis, interaction energy $|\Phi_n(t)\rangle$, will vanish due to the time average of the factor $e^{i(\Lambda_n(t)-\Lambda_{n^\prime}(t))/\hbar}$. 

In the case that $\{|\Phi_{0n}(t)\rangle\}$ are the simultaneous eigenstates of $\hat{h}_\phi+\hat{h}_\varepsilon$ and $\hat{h}_{int}$ (both assumed to be independent of time), such a simplefied approach is still correct (the ideal measurement case). 
It, however, is necessary to discuss the reason that the interaction energy is privileged compared with the system's exact internal energy.

\subsection{Fluctuation and suppression of unitary evolution}
We here study the transitional behavior of macroscopic quantum systems for a intermediate time interval. 
The important and non-obvious case is that the interaction Hamiltonian is not commutative with the system's Hamiltonian in its strict sense. 
We see below that the continuous spectrum of the interaction Hamiltonian, $\hat{h}_{int}$,  recovers even in our approach.
Therefore it is necessary to consider the interaction energy approach for an arbitrary non-perturbed state: the state evolves into, $|\Phi(t)\rangle=|\Phi_0(t)\rangle e^{-i\Lambda_\Phi(t)/\hbar}$, which obviously satisfies 
\begin{eqnarray}\label{eq:mean-field}
[i\hbar\partial_t-\hat{h}_\phi-\hat{h}_\varepsilon-\langle\Phi(t)|\hat{h}_{int}|\Phi(t)\rangle]|\Phi(t)\rangle=0, 
\end{eqnarray}
where $\Lambda_\Phi=\int^t_{t_0} \langle\Phi(t)|\hat{h}_{int}|\Phi(t)\rangle dt.$
From the fact that this equation has the non-linearlity depending on $|\Phi(t)\rangle$, 
we reconsider the expansion of the total system with the basis $\{|\Phi_n(t)\rangle\}.$
The interaction $\hat{h}_{int}$ requires us to extent the state space (we should remember the  interaction Hamiltonian, $\hat{h}_{int}$, and then the exact Hamiltonian have continuous spectra).

We then consider carefully the superposition of the product states paying attention to both the  linearity of the equation (\ref{eq:whole}) and the time dependence of $\Lambda(t)$.
Two non-orthogonal initial states, $\langle\Phi_0(t_0)|\Phi_0^\prime(t_0)\rangle\ne 0,$  evolve into the states, $|\Phi(t)\rangle=|\Phi_0(t)\rangle e^{-i\Lambda_\Phi(t)/\hbar}$ and $|\Phi^\prime(t)\rangle=|\Phi_0^\prime(t)\rangle e^{-i\Lambda_{\Phi^\prime(t)}/\hbar}$  with different $\Lambda$'s in general.
Therefore these must be treated as the {\bf linearly independent solutions} of Eq. (\ref{eq:whole}) because the time dependence of $\Lambda_\Phi-\Lambda_{\Phi^\prime}$ leads the orthogonality relation, 
\begin{eqnarray}
\int^\infty_{-\infty}\langle\Phi(t)|\Phi^\prime(t)\rangle dt=\langle\Phi_0(t)|\Phi_0^\prime(t)\rangle\int^\infty_{-\infty}e^{i(\Lambda_\Phi-\Lambda_{\Phi^\prime})/\hbar}dt=0. 
\end{eqnarray}
(We used the fact that the coefficient $\langle\Phi_0(t)|\Phi_0^\prime(t)\rangle$ is  independent of time as immediately represented in the eigenstates of the non-perturbed  Hamiltonian, $\{|\Phi_{0\epsilon}\rangle\}$. The expansions, $|\Phi_0(t)\rangle=\sum_\epsilon C_\epsilon|\Phi_{0\epsilon}\rangle e^{-i\epsilon t/\hbar}$ and $|\Phi_0^\prime(t)\rangle=\sum_\epsilon C^\prime_\epsilon|\Phi_{0\epsilon}\rangle e^{-i\epsilon t/\hbar}$, lead $\langle\Phi_0(t)|\Phi_0^\prime(t)\rangle=\sum_\epsilon C^\ast_\epsilon C_\epsilon^\prime$.
Moreover we here assume that $\Lambda$'s monotonously increase because of the continuous interaction between the system and its environment after $t_0$. )
Therefore the linear combination of the states of the original system, $|\Phi_{0\nu}(t)\rangle=\sum_nC_{\nu n}|\Phi_{0n}\rangle$, is necessary in order to expand the state of the system when the interaction exists, where index $\nu$ corresponds different value of $\Lambda$.
\begin{eqnarray}\label{eq:expansion3}
|\Phi(t)\rangle\simeq\sum_\nu C_\nu(t_0)|\Phi_{0\nu}(t)\rangle e^{-i\Lambda_\nu(t)/\hbar}.
\end{eqnarray}
It is important that the continuous spectrum of the exact Hamiltonian $\hat{h}_\phi+\hat{h}_\varepsilon+\hat{h}_{int}$ recovers even in such the lowest-order approach. 

In the case of macroscopic systems, however, the destructive interference (mathematically the saddle point approximation), is expected to lead the suppression of the fluctuation again:
\begin{eqnarray}\label{eq:expansion4}
|\Phi(t)\rangle&&\simeq\sum_\nu C_\nu(t_0)|\Phi_{0\nu}(t)\rangle e^{-i\Lambda_\nu(t)/\hbar}\\
&&\to\sum_{\nu_c} C_{\nu_c}(t_0)|\Phi_{0\nu_c}(t)\rangle e^{-i\Lambda_{\nu_c}(t)/\hbar},
\end{eqnarray}
where the states, $\{\Phi_{0\nu_c}\rangle\}$, give extreme values of $\Lambda$. 

\vspace{1cm}

\section{Two-state system embedded in a Coulomb potential}
Although we intend to expand the state of the system by the eigenstate of the interaction Hamiltonian, we find below that the number of the states composing the mixture of states is only two. 
The destructive interference suppresses the fluctuation as mentioned in \S 2.3. 

In the previous section we found that the lowest-order (particular) solution is constructed with the solutions of (\ref{eq:subsystem1}) and (\ref{eq:subsystem2}) in the product form as, 
\begin{eqnarray}\label{eq:action}
|\Phi(t)\rangle=|\Phi_0(t)\rangle e^{-i\Lambda(t)/\hbar}=|\phi(t)\rangle|\varepsilon(t)\rangle e^{-i\Lambda(t)/\hbar},
\end{eqnarray}
and here the action, $\Lambda(t)$, is the time integrated interaction energy: 
\begin{eqnarray}
\Lambda(t)=\int^t_{t_0}\langle \phi(t)|\hat{V}|\phi(t)\rangle dt
=\int^t_{t_0}\langle \Phi(t)|\hat{h}_{int}|\Phi(t)\rangle dt,
\end{eqnarray}
where $\hat{V}=\langle \varepsilon(t)|\hat{h}_{int}|\varepsilon(t)\rangle$ is the `external' field.

Let us study further the case that the macroscopic system, $|\phi(t_0)\rangle$, is considered as a two-state system: we define $|\phi_\uparrow(t_0)\rangle$ and $|\phi_\downarrow(t_0)\rangle$ as the two normalized eigenstates of $\hat{h}_\phi$ being orthogonal each other.
Namely we assume that the state of a macroscopic system is represented by the linear combination of only two states and the other states can be neglected.
As the result, the interaction energy can also have two extrema as below. 
Moreover we assume the interaction energy is independent of the state of the environment, $|\varepsilon(t)\rangle$, just for simplicity.

\subsection{Nearly-commutative case}
Using these states, we define the following linear combination in order to find out one particular solution of Eq. (\ref{eq:whole}): 
\begin{eqnarray}
|\phi_\theta(t_0)\rangle=\cos\theta|\phi_\uparrow(t_0)\rangle+\sin\theta|\phi_\downarrow(t_0)\rangle.
\end{eqnarray}
If the interaction Hamiltonian represented by the energy eigenstates, $|\phi_\uparrow(t_0)\rangle$ and $|\phi_\downarrow(t_0)\rangle$, is almost diagonal, the time integrated interaction energy has the simple form as, 
\begin{eqnarray}
\Lambda_\theta(t)&=&\cos^2\theta\int^t_{t_0}\langle \phi_\uparrow(t)|\hat{V}|\phi_\uparrow(t)\rangle dt+\sin^2\theta\int^t_{t_0}\langle  \phi_\downarrow(t)|\hat{V}|\phi_\downarrow(t)\rangle dt\\
&\equiv&\cos^2\theta\Lambda_\uparrow(t)+\sin^2\theta\Lambda_\downarrow(t).
\end{eqnarray}
We can see here that $\theta$ is the index corresponding to the continuous spectrum of $\hat{V}$.

Therefore now the initial states simply evolves as, 
\begin{eqnarray}
|\Phi_\theta(t_0)\rangle=|\phi_\theta(t_0)\rangle|\varepsilon(t_0)\rangle \to |\Phi_\theta(t)\rangle=|\phi_\theta(t)\rangle|\varepsilon(t)\rangle e^{-i\Lambda_\theta(t)/\hbar}, 
\end{eqnarray}
where we can see that the environment only changes its phase from $|\varepsilon(t_0)\rangle$ to $|\varepsilon_\theta(t)\rangle=|\varepsilon(t)\rangle e^{-i\Lambda_\theta(t)/\hbar}$.

We here notice again that $|\Phi_\theta(t)\rangle$ depends on the coefficients $\cos^2\theta$ and $\sin^2\theta$ through the action. 
It is clear that $\Lambda_\uparrow(t)\ne\Lambda_\downarrow(t)$ reproduces the orthogonality relation $\langle \varepsilon_\theta(t)|\varepsilon_{\theta^\prime}(t)\rangle=e^{i(\Lambda_\theta(t)-\Lambda_{\theta^\prime}(t))/\hbar}\to\delta_{\theta,\theta^\prime}$ for {\bf time average}.
At the same time, the corresponding states of the total system, $|\Phi_\theta(t)\rangle$ and $|\Phi_{\theta^\prime}(t)\rangle$, are the linearly independent solutions of Eq. (\ref{eq:whole}) for $\Lambda_\theta(t)\ne\Lambda_\theta^\prime(t)$: 
\begin{eqnarray}\label{eq:orthogonality}
\int^\infty_{t_0}\langle \Phi_\theta(t)|\Phi_{\theta^\prime}(t)\rangle dt=\cos(\theta-\theta^\prime)\int^\infty_{t_0}dt e^{-i(\Lambda_\theta(t)-\Lambda_{\theta^\prime}(t))/\hbar}\propto\delta_{\theta,\theta^\prime}.
\end{eqnarray}
If we assume the constant monitoring of macroscopic object by its environment, i.e., $\Lambda(t)=\lambda (t-t_0)$, we estimate that the orthogonality is achieved in very short time, $\tau\simeq \hbar/\lambda$, where the interaction energy, $\lambda=\langle\phi(t)|\hat{V}|\phi(t)\rangle$, is a macroscopic quantity. 

Therefore the interaction between subsystems fractionates the expansion of the total system  through $\Lambda,$ and the linear combination of $|\Phi_\theta(t)\rangle$'s, is {\bf necessarily} required. 
An arbitrary states at $t_0$ is expressed as $|\phi(t_0)\rangle=e^{i\alpha}\cos\theta|\phi_\uparrow(t_0)\rangle+e^{i\beta}\sin\theta|\phi_\downarrow(t_0)\rangle$, but $\Lambda$ is independent of these phase factors, $\alpha$ and $\beta$, in the present model.
Then, in the mean-field level, the (general) solution of Eq. (\ref{eq:whole}) is represented by, 
\begin{eqnarray}\label{eq:Phit}
|\Phi(t)\rangle&&=\sum_{0\le\theta\le\pi/2} C_\theta|\Phi_\theta(t)\rangle\\
&&=\sum_{0\le\theta\le\pi/2} C_\theta |\phi_\theta(t)\rangle|\varepsilon(t)\rangle e^{-i\Lambda_\theta(t)/\hbar}\\
&&=\sum_{0\le\theta\le\pi/2} C_\theta(\cos\theta|\phi_\uparrow(t)\rangle+\sin\theta|\phi_\downarrow(t)\rangle)|\varepsilon_\theta(t)\rangle,
\end{eqnarray} 
where $\sum_\theta|C_\theta|^2=1$. 
(For a sufficiently long time, although the angle $\theta$ is continuous, we use the notation $\sum_\theta$ just for simplicity. 
The number of the basis, being orthogonal each other, depend on the interacting time $t-t_0.$ )
The initial states are reorganized into new groups labeled by $\theta$. 

It is natural that the initial states are fractionized when the system is embedded in the ordinary potential with continuous spectrum. 
We notice that only giving the initial conditions $\alpha$ and $\beta$ for the subsytem $|\phi(t)\rangle=\alpha|\phi_\uparrow(t)\rangle+\beta|\phi_\downarrow(t)\rangle$ is generally insufficient to determine the evolution of the whole system $|\Phi(t)\rangle$ with no ambiguity 
because the initial state, $|\Phi(t_0)\rangle=|\Phi_0(t_0)\rangle=\left[(\sum_\theta C_\theta \cos\theta)|\phi_\uparrow(t_0)\rangle+(\sum_\theta C_\theta \sin\theta)|\phi_\downarrow(t_0)\rangle\right]|\varepsilon(t)\rangle$, has a kind of degeneracy for $\theta$.
In other words, in appearance, different time evolutions can emerge from an identical initial condition, $|\phi(t)\rangle=\alpha|\phi_\uparrow(t)\rangle+\beta|\phi_\downarrow(t)\rangle$.

\bigskip
{\bf Suppression of fluctuation:}
For $t>t_0,$ the interaction between the macroscopic systems, $|\phi(t)\rangle$ and $|\varepsilon(t)\rangle,$ naturally makes the action $\Lambda(t)$ large. 
Under the condition $\Lambda_\theta(t)\gg\hbar$, we can adopt the {\bf saddle point  approximation} on the solution: the contribution of the terms that satisfy $\displaystyle \frac{\delta}{\delta\theta}\Lambda_\theta(t)=0$ only survive in equation (\ref{eq:Phit}),
\begin{eqnarray}
|\Phi(t)\rangle&\sim& \tilde{C}_0|\phi_\uparrow(t)\rangle|\varepsilon(t)\rangle e^{-i\Lambda_\uparrow(t)/\hbar}+
\tilde{C}_{\pi/2}|\phi_\downarrow(t)\rangle|\varepsilon(t)\rangle e^{-i\Lambda_\downarrow(t)/\hbar}\\
&=&\tilde{C}_0|\phi_\uparrow(t)\rangle|\varepsilon_\uparrow(t)\rangle+
\tilde{C}_{\pi/2}|\phi_\downarrow(t)\rangle|\varepsilon_\downarrow(t)\rangle,\label{eq:approximatevectors}
\end{eqnarray}
where $\tilde{C}_{0, \pi/2}=C_{0, \pi/2}\sqrt{\frac{2\pi i}{\Lambda_{0, \pi/2}^{\prime\prime}(t)/\hbar}}=C_{0, \pi/2}\sqrt{\frac{\pm\pi i\hbar}{\Lambda_\uparrow(t)-\Lambda_\downarrow(t)}}$.
Namely, the two-state system recovers again due to the destructive interference.
The resulting states, however, correspond to the extreme values of interaction energy.

Now we can conclude that the condition $\Lambda_\theta(t)\gg\hbar$ (more explicitly the variation of $\Lambda_\theta(t)/\hbar$ being large and also assumed $\Lambda_\uparrow(t)\ne\Lambda_\downarrow(t)$) leads that only one basis, $\{|\Phi_\theta(t)\rangle=|\phi_\theta(t)\rangle|\varepsilon_\theta(t)\rangle\}$, is privileged to express the total system $|\Phi(t)\rangle$:
This basis consists of the $|\Phi_\theta(t)\rangle$'s, which give extremum to the action, $\Lambda_\theta(t)$.
We have demonstrated the emergence of pointer basis in terms of state vectors. 

{\bf Density matrix:} In the present context, the reduced density matrix of the subsystem $\rho_{SA}={\rm Tr}_\varepsilon|\Phi(t)\rangle\langle\Phi(t)|$ has the off-diagonal terms with the factor $r(t)=\langle\varepsilon_\uparrow(t)|\varepsilon_\downarrow(t)\rangle=e^{i(\Lambda_\uparrow(t)-\Lambda_\downarrow(t))/\hbar}.$ 
These will vanish for the time average.

\subsection{Non-commutative case}
When the linear combination of the energy eigenstates, such as $|\phi_\pm(t_0)\rangle\equiv|\phi_\downarrow(t_0)\rangle/\sqrt{2}\pm|\phi_\downarrow(t_0)\rangle/\sqrt{2}$, give extreme values of interaction energy at the specific time ($t_0$), 
we can say that the interaction Hamiltonian, $\hat{V}$, and the system's Hamiltonian, $\hat{h}$, are strongly non-commutative. 
In this case, although the expected privileged-states are $|\phi_+(t)\rangle$ and $|\phi_-(t)\rangle$, 
these states strongly depend on time by the system's Hamiltonian. 
Therefore the interaction energy and the off-diagonal elements, $\langle \phi_\pm(t)|\hat{V}|\phi_\mp(t)\rangle$, have oscillational terms.
Assuming a short interval and $\hat{V}|\phi_\pm(t_0)\rangle=V_\pm|\phi_\pm(t_0)\rangle$, 
\begin{eqnarray}
\langle\phi_+(t)|\hat{V}|\phi_-(t)\rangle&&\simeq\frac{i(t-t_0)}{\hbar}\langle\phi_+(t_0)|[\hat{h}, \hat{V}]|\phi_-(t_0)\rangle\\
&&=\frac{i(t-t_0)}{\hbar}(V_--V_+)\langle\phi_+(t_0)|\hat{h}|\phi_-(t_0)\rangle\\
&&=\frac{i(t-t_0)}{2\hbar}(V_--V_+)(\varepsilon_\uparrow-\varepsilon_\downarrow).
\end{eqnarray}
We can easily see that our assumption of diagonality, $\langle \phi_\pm(t)|\hat{V}|\phi_\mp(t)\rangle\simeq 0$, (which derives eq. (\ref{eq:diagonal})) 
cannot be satisfied within the short time, $\tau=\frac{2\hbar}{|\varepsilon_\uparrow-\varepsilon_\downarrow|}\ll\frac{\hbar}{|V_+-V_-|}$, in the case of weak perturbation. 
It is necessary for the quantum decoherence of a two state system that the interaction Hamiltonian 
and the system's Hamiltonian are sufficiently commutative.

\bigskip

On the other hand, if the initial state consists of the number of energy eigenstates, the process of the decoherence will be more time-evolutional. 
(Naively we can image this by replacing $\tau=\frac{2\hbar}{|\varepsilon_\uparrow-\varepsilon_\downarrow|}$ with $\tau=\frac{N\hbar}{\Delta \varepsilon}=O(\sqrt{N})$. )
In the next section, we consider the case of the localization of macroscopic object, where the eigenstates of the interaction Hamiltonian can be continuously generated.

\vspace{1cm}

\section{Localization of macroscopic object}
Quantum mechanics can give the precise knowledge about the internal state of a matter.
On the other hand, it seems to fail in reproducing the behavior of the center of mass of a macroscopic object.
Let us start from comfirming this fact below. 
In this section, we use the wave function in terms of the coordinate representation. 

In many cases, the Hamiltonian for the center of mass coordinats is separated from the internal degrees of freedom, $\hat{h}=\hat{h}_{\bf r}+\hat{h}_{\bf R}$, where ${\bf r}$ denotes the internal degrees of freedom and ${\bf R}$ denotes the center of mass. 
Then the state of a macroscopic object is, 
\begin{eqnarray}
\phi({\bf r}, {\bf R}, t)=\phi_{cm}({\bf R}, t)\phi_{internal}({\bf r}, t).
\end{eqnarray}

For the isolated case, $\hat{h}_{\bf R}=-\hbar^2\nabla^2_{\bf R}/2M$, the wave function is represented as 
\begin{eqnarray}
\phi_{cm}({\bf R}, t)=\sum_{\bf k}a_{\bf k}e^{i{\bf k}\cdot{\bf R}-i\varepsilon_{\bf k}t/\hbar}.
\end{eqnarray}
Of course, the long wave limit of the plane wave state has the lowest energy, 
\begin{eqnarray}
\phi_{cm}^{lowest}({\bf R},t)=\lim_{{\bf k}\to 0}a_{\bf k}e^{i{\bf k}\cdot{\bf R}-i\varepsilon_{\bf k}t/\hbar}.
\end{eqnarray}
We, however, cannot observe the plane-wave state of a object and the localization of its center of mass is always realized. 
We cannot naively attribute the reason to the external potential: 
Although there is the case that an atractive or a repulsive weak perturbation can be added on the Hamiltonian $\hat{h}_{cm}$, the localization of the macroscopic object usually costs higher energy.

\bigskip
Several important results are known about the role of the quantum decoherence on the localization of macroscopic object.\cite{Joos1985}
As we have also mentioned above, it is known that the pointer states which diagonalize the interaction Hamiltonian are privileged. 
Therefore, in the case that a quantum system interacts with its environment through the center of mass coordinate, the pointer states will be the eigenstates of the interaction Hamiltonian, where these states are localized in the coordinate space. 

Moreover, Joos and Zeh\cite{Joos1985} demonstrated that the quantum states with the different position exponentially lost the quantum coherence by scattering effect its environment. 
The reduced density matrix plays important role to reproduce the mixture of states.
Although we agree with these researches on the crucial role of the interaction with the environment, we focus especially on the interaction energy below.

\bigskip

As seen in \S 2, in appearance, the non-lineality is inherent in the mean-field approximation, and its dynamics produces the orthogonality in terms of time average.
Finally the states which give extrema of the interaction energy survive due to the destructive interference. 
The interaction Hamiltonian is naturally diagonal in the basis consisting of these states because the saddle point condition $\delta\Lambda=0$ always selects out the approximate eigen states of the interaction Hamiltonian from the all of the solutions $\{\Phi_\nu\}$. 
(We have assumed that the initial state contains sufficient number of states when it is represented by the wave number, ${\bf k}.$ 
Since the emerging states consist of the linear combination of these initial states, as is seen in previous section, this assumption is important so that the diagonality relation is satisfied by the saddle point condition $\delta\Lambda=0$. )
Then, the mixture of states are realized, $\Phi\simeq\sum_{\nu_c}a_{\nu_c}\Phi_{\nu_c}e^{-i\Lambda_{\nu_c}/\hbar}$.

\bigskip

We here consider the case that the quantum system interacts with its environment, $\Psi_{env}(t)$, through the center of mass coordinate, ${\bf R}$. 
In this case, after the time interval $t_{dec}$, the emerging pointer states will be the approximate eigenstates of the external field $V({\bf R})$: 
\begin{eqnarray}
\hspace{-0.5em}\Phi(t)
&&\hspace{-1em}=e^{-i(\hat{h}_{\bf R}+\hat{h}_{\bf r}+\hat{h}_{int}+\hat{h}_{env})(t-t_0)/\hbar}\phi_{cm}({\bf R}, t_0)\phi_{internal}({\bf r}, t_0)\Psi_{env}(t_0)\\
&&\hspace{-2.5em}=e^{-i(\hat{h}_{\bf R}+\hat{h}_{int})(t-t_0)/\hbar}\phi_{cm}({\bf R}, t_0)\phi_{internal}({\bf r}, t)\Psi_{env}(t)\\
&&\hspace{-2.5em}\simeq e^{-i\hat{h}_{\bf R}(t-t_0)/\hbar}\sum_{\nu}C_\nu \phi_{cm,\nu}({\bf R}, t_0)e^{-i\Lambda_\nu/\hbar}\phi_{internal}({\bf r}, t)\Psi_{env}(t)\\
&&\hspace{-2.5em}=e^{-i\hat{h}_{\bf R}(t-t_0)/\hbar}\sum_{\nu}C_\nu [\sum_{\bf k}a_{{\bf k},\nu}e^{i{\bf k}\cdot{\bf R}}]e^{-i\Lambda_\nu/\hbar}\phi_{internal}({\bf r}, t)\Psi_{env}(t)\\
&&\hspace{-2.5em}\simeq e^{-i\hat{h}_{\bf R}(t-t_0)/\hbar}\sum_{\nu_c}C_{\nu_c} [a_{\nu_c}\sum_{\bf k}e^{i{\bf k}\cdot({\bf R}-{\bf R}_{\nu_c})}]e^{-iV({\bf R}_{\nu_c})(t-t_0)/\hbar}\phi_{internal}({\bf r}, t)\Psi_{env}(t)\\
&&\hspace{-2.5em}=\sum_{\nu_c}\tilde{C}_{\nu_c}[\sum_{\bf k}e^{i{\bf k}\cdot({\bf R}-{\bf R}_{\nu_c})-i\varepsilon_{\bf k}(t-t_0)/\hbar}]e^{-iV({\bf R}_{\nu_c})(t-t_0)/\hbar}\phi_{internal}({\bf r}, t)\Psi_{env}(t).
\end{eqnarray}
The last equation represents just the Huygens' wave propagation except for the oscillation factor $e^{-iV({\bf R}_{\nu_c})(t-t_0)/\hbar}$ which originates from the interaction energy. 
In the third equation, referring to the interaction energy, the state fractionized by the perturbation are grouped by the index $\nu$, assuming that there is no degeneracy.   
In the fourth equation, the state, $\phi_{cm,\nu}(t_0)$, is represented by the eigenstates of $\hat{h}_{\bf R}$, just for easy understanding.
In the fifth equation, we have used the saddle point approximation, where $\delta \Lambda/\delta a^\ast_{\bf k}=0$ leads $a_{\bf k}\simeq a e^{-i{\bf k}\cdot{\bf R}_{\nu_c}}.$ 
(Here, $\Lambda$ is the functional of $a_{\bf k}$ and the Lagrange multiplyer from the norm condition is necessary.)
This means the localization of the states, $|\phi_{cm,\nu_c}|^2\simeq\delta({\bf R}-{\bf R}_{\nu_c})$. 
(We can directly derive this by considering $\delta\Lambda/\delta\phi_{cm}=0$.)
Then we can calculate the interaction energy as, $E_{int}=\langle\Phi|\hat{h}_{int}|\Phi\rangle=\int\phi_{cm}^\ast V({\bf R})\phi_{cm} d{\bf R}\simeq V({\bf R}_{\nu_c})$.

\bigskip
Of course, since the localized states are not the eigenstates of the Hamiltonian, $\hat{h}_{\bf R}=-\hbar^2\nabla^2/2M$, these states may spread after the specific time $t_{coh}$. 
For a macroscopic object, however, we can expect that $t_{dec}\ll t_{coh}$ because of its large mass and each state will repeatedly branch into localized states, where 
\begin{eqnarray}
t_{dec}\equiv\frac{\hbar}{\langle{\bf R}|V|{\bf R}\rangle-\langle{\bf R}^\prime|V|{\bf R}^\prime\rangle}, \hspace{0.5cm}
t_{coh}\equiv\frac{\hbar}{\langle{\bf R}|\hat{h}_{\bf R}|{\bf R}^\prime\rangle}. 
\end{eqnarray}
This relation is important also in the previous calculation steps.
For the given time interval $T$, the resolution $\Delta{\bf R}={\bf R}-{\bf R}^\prime$ is defined so that the relation, $t_{dec}\ll T,$ holds.

Since the unitary evolution due to the Hamiltonian, $\hat{h}_{\bf R}$, 
holds in each branch, classical mechanics of material point is reproduced in each branch as Ehrenfest's theorem shows.

\vspace{1cm}

\section{Discussion}

\noindent{\bf Mixture of states and reality}\\
From the early days of quantum mechanics, it is recognized that an individual event observed in the microscopic world often seems to be part of whole that is predected by quantum mechanics. 
Even if we successfully accomplish the decoherence program, in which the emergence of the mixture of states is clearly shown, the gap between the objective reality and the subjective description will be still left\cite{Schlosshauer2005}. 
This gap, however, seems to narrow if our macroscopic reality irreversibly and repeatedly splits off into branches, where each branch consists of entangled quantum systems. 

For instance, our conciousness may be identified as the specific quantum state, $\Phi_{\nu_c}$, of the nerves.
Under this idea, for the observer in a specific branch, the interaction inducing mixture does not exist effectively:
It does not contribute to the `physical' entropy since the other branches represent just the possible (but invisible) phenomena. 
In this sense the emergence of the entanglement is relevant to getting information.\footnote{
Since the present discussion is pure speculation, the author recommend that the reader refers other precise researches. For example, see the recent papers in the references.} 
(On the other hand, for physical entropy, the concept of probability or coarse-graining is not essential when we adopt von Neumann's derivation of the thermal mixture.)

\bigskip
\noindent{\bf Dissipative feature}\\
A macroscopic system is usually considered to have a finite entropy. 
This means that this system is expressed as the mixture of states: 
In fact, von Neumann demonstrated that a density matrix of usual (pure) macroscopic system can be identified to the micro-canonical density matrix if we consider the long time average.
The irreversible feature in our experience, however, can't be deriven since the time evolution of an isolated quantum system is described by unitary operators.
(So Von Neumann suggested the relevance of the measurement process on this subject in the footnote of \cite{Neumann1929}, p5)

In this study, we have found that a `visible' degrees of freedom, such as the center of mass, appears in front of us 
through the entanglement between the system and its environment. 
In other words, a quantum system emerges into each environment ($\Psi_{\nu_c}$) as an approximate eigenstate (pure state $\phi_{\nu_c}$) of the interaction Hamiltonian. 
Therefore such a visible degree of freedom of a certain physical quantity has zero entropy. 
The quantity belonging to this degree of freedom, however, will naturally spread over into other invisible ones (in the above context, denoted by $\phi_{internal}({\bf r},t)$ or $\Psi_{env}(t)$) due to unitary evolution.
Let us remember that the states consisting of these invisible degrees of freedom are in mixtures of states (and have finite physical entropy) when it is represented by the energy eigenstate.
Resultingly the dissipative feature of this world is suggested if it is allowed to assume that each entangled states represents an individual reality.
It, however, is the future task to reproduce such a process with the use of general and concrete models.

\end{document}